\documentstyle[aps,prl,epsf]{revtex}

\begin{document}

\title{\bf \LARGE {\it Appendix to:}
Energy Density Functional Approach to Superfluid Nuclei,
 nucl-th/0210047 }
\vspace{0.75cm}
\author{Yongle Yu and Aurel Bulgac}
\vspace{0.50cm}
\address{Department of Physics, University of
Washington, Seattle, WA 98195--1560, USA}

\maketitle

\vspace{0.75cm}
\today
\vspace{0.75cm}

\begin{abstract}

This report is not meant to be submitted for publication at the
present time and it is simply a supplement to our work \cite{sedf}.  We have
collected together all our results for one--nucleon and two--nucleon
separation energies for several isotope and isotone chains and compare
them to the values extracted from the 1995 Audi and Wapstra table of
recommended masses \cite{audi}. Where it was possible we have compared
our results to the results of Fayans {\it et al.} \cite{fayans} and Goriely
{\it et al.} \cite{goriely}. We also present results for the charge radii
and compare them to experiment and results of  Goriely {\it et al}. 

\end{abstract}

\draft

The entire formalism has been described in Ref. \cite{sedf} and
earlier references cited therein. Since the best agreement between our
energy density functional for superfluid nuclei and experimental
binding energies  was obtained when we use for the normal part of this
functional Fayans' FaNDF$^0$ and the bare pairing coupling constant $g
= -200\; MeV\; fm^{-3}$, we limit the results presented here to this
case only. For the experimental binding energies we have used the
recommended values extracted from Ref. \cite{audi}. The results of
Fayans {\it et al.}  have been obtained by digitizing their published
figures and for that reason inaccuracies are likely. The reader has to
keep in mind also that Fayans {\it et al.} \cite{fayans} did not use a
universal pairing energy and have renormalized the strength of their
interaction by a factor of 1.05 in case of tin isotopes and by a
factor of 1.35 in case of calcium isotopes. Since these authors have
also published results for one--nucleon separation energies without
this renormalization of the pairing interaction, we have extracted the
unrenormalized results in case of $S_n$ for Ca isotopes. Fayans {\it
et al.} \cite{fayans} have performed calculations only for these three
isotope chains. The results of Goriely {\it et al.}  \cite{goriely}
have been extracted from their website. Goriely {\it et al.}
\cite{goriely} use separate pairing strengths for even proton, odd
proton, even neutron and odd neutron systems. In all our calculations
we use the same pairing strength for both proton and neutron systems
and we did not change its value as a function of the atomic number. We
have treated all nuclei as spherical. It is well known that many
nickel isotopes are deformed, see e.g. Goriely {\it et al.}
\cite{goriely}, and we suspect that some discrepancies between our
predictions and experimental data can be ascribed to
deformation. Since it is known that there is a significant amount of
pairing among protons in zirconium isotopes\cite{negele}, pairing has
been allowed in both proton and neutron subsystems, but only with the
couplings $g=g_0+g_1\ne 0$ and $g^\prime = g_0-g_1\equiv 0$, see
Ref. \cite{sedf} for details. Note also that the heavier zirconium
isotopes are also deformed according to Goriely {\it et al.}

In the last five figures we present the charge radii for Ca, Ni, Zr,
Sn and Pb isotopes. For the proton charge radius we have used
$r_p=0.87\; fm$ according to Ref. \cite{pdg}. Note that Goriely {\it
et al.} use instead an older value of $r_p=0.80\; fm$. The
experimental values for the nuclear charge radii are from Nadjakov
{\it et al.} \cite{rch} and were very kindly provided to us by
J. Michael Pearson, to whom we are very thankful. In our
self--consistent calculations the protons were treated as point
particles when evaluating the nuclear Coulomb energy. Fayans made a
strong point in private discussions with us for using the charge
distribution with the finite formfactors for both protons and
neutrons. A finite proton size leads to a smaller Coulomb energy and a
less repulsive Coulomb potential for protons, which thus leads to
slightly larger charge radii and very likely to a better overall
agreement with experiment in our case. We have also neglected the
Coulomb exchange and correlation energies and a possible charge
symmetry contribution to the total energy as well, see
Refs. \cite{sedf,fayans} for additional references and discussions of
these terms. Except for the cases when deformation is likely to affect
significantly the values of the charge radii, the agreement between
our theoretical predictions and experiment is about of the same
quality as in case of Goriely {\it el.} All together we present
results for 212 nuclei.

\begin{figure}[tbh]
\begin{center}
\epsfxsize=10.0cm
\centerline{\epsffile{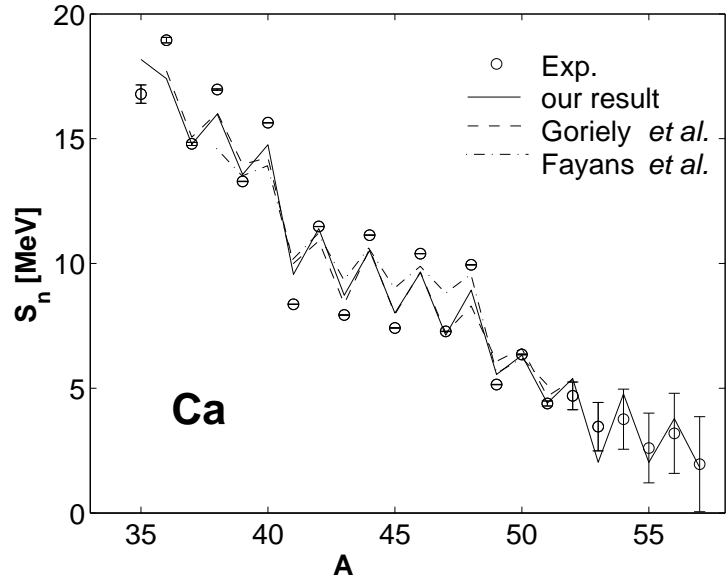}}
\end{center}
\caption{ One--neutron separation energies for calcium isotopes.  The
results of Fayans {\it et al.}  presented here are for unrenormalized
pairing strength.}

\end{figure}

\begin{figure}[tbh]
\begin{center}
\epsfxsize=10.0cm
\centerline{\epsffile{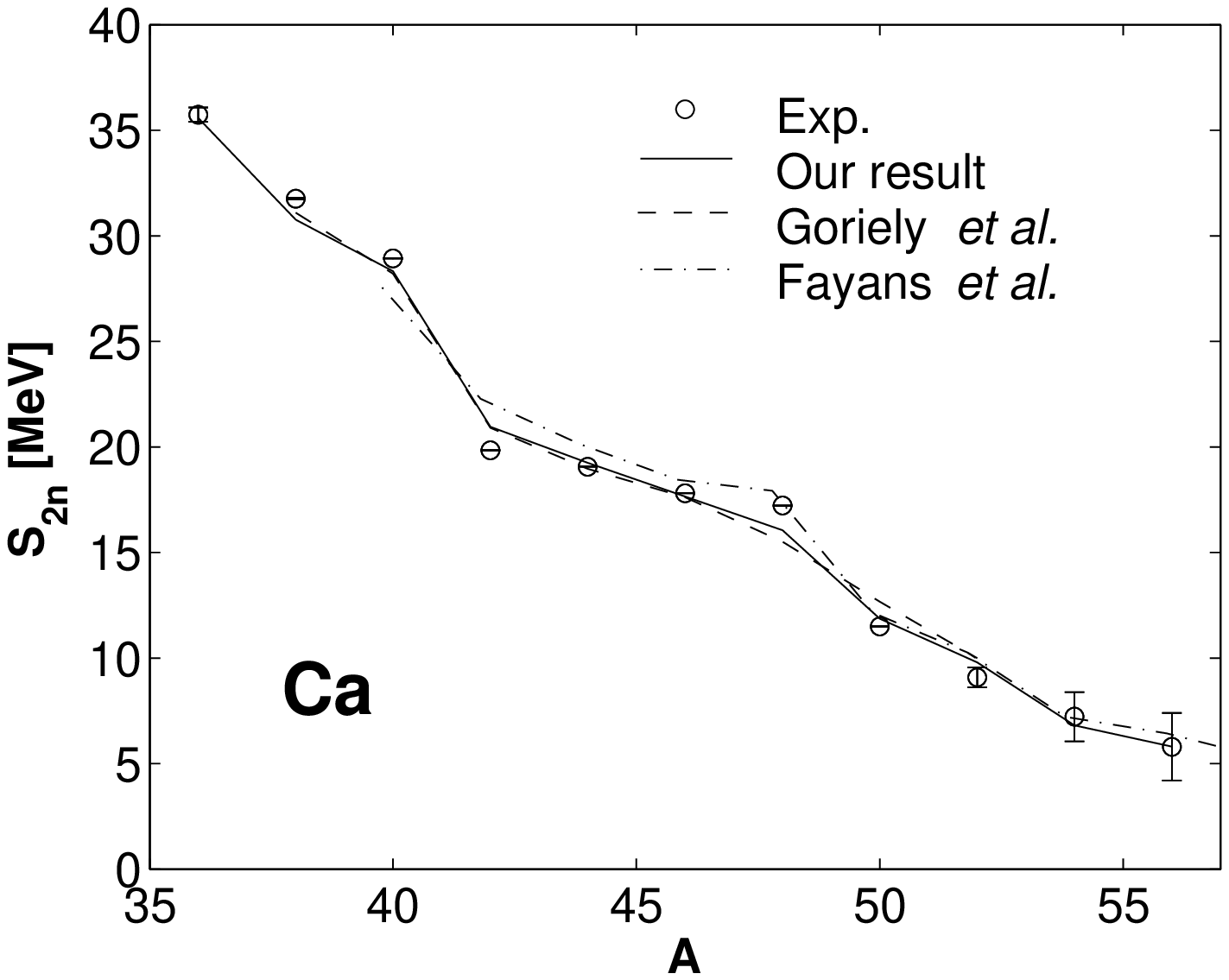}}
\end{center}
\caption{Two--neutron separation energies for calcium isotopes. The
results of Fayans {\it et al.} presented here are for a renormalized
pairing strength with a factor of 1.35. }
\end{figure}

\begin{figure}[tbh]
\begin{center}
\epsfxsize=10.0cm
\centerline{\epsffile{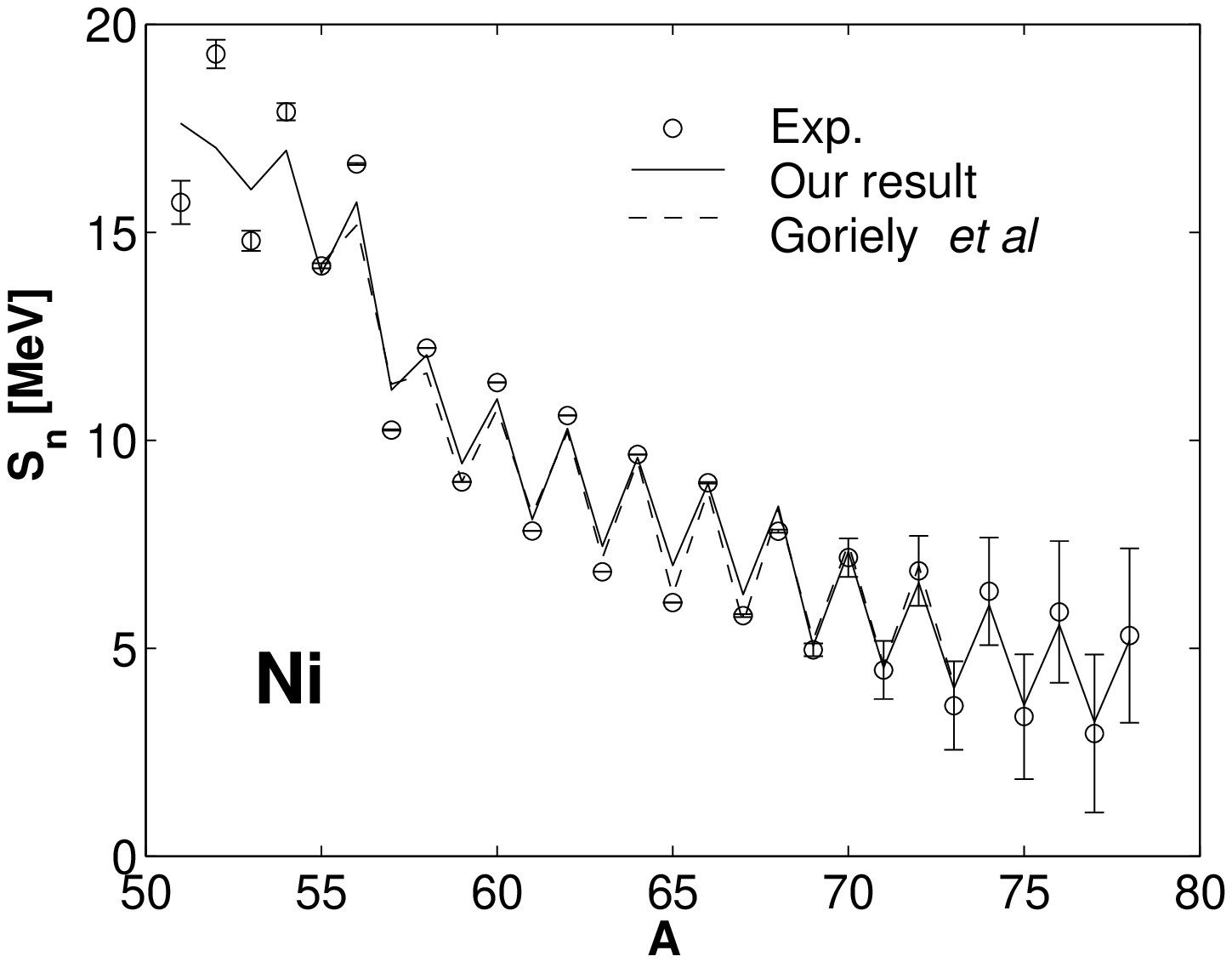}}
\end{center}
\caption{One--neutron separation energies for nickel isotopes. 
Note that many nickel isotopes are deformed and we have treated 
them as spherical.}
\end{figure}

\begin{figure}[tbh]
\begin{center}
\epsfxsize=10.0cm
\centerline{\epsffile{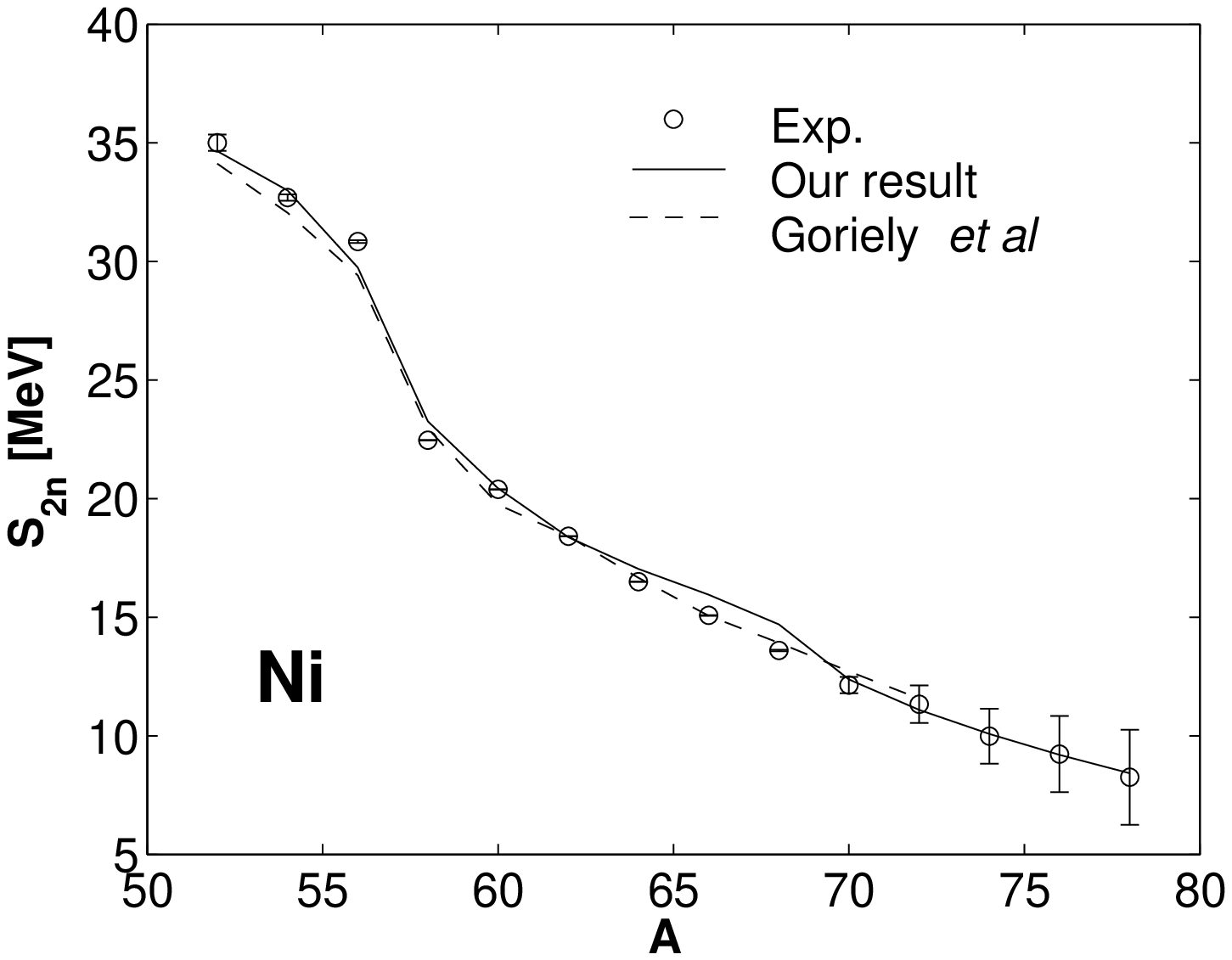}}
\end{center}
\caption{Two--neutron separation energies for nickel isotopes. 
Note that many nickel isotopes are deformed and we have treated 
them as spherical.}
\end{figure}

\begin{figure}[tbh]
\begin{center}
\epsfxsize=10.0cm
\centerline{\epsffile{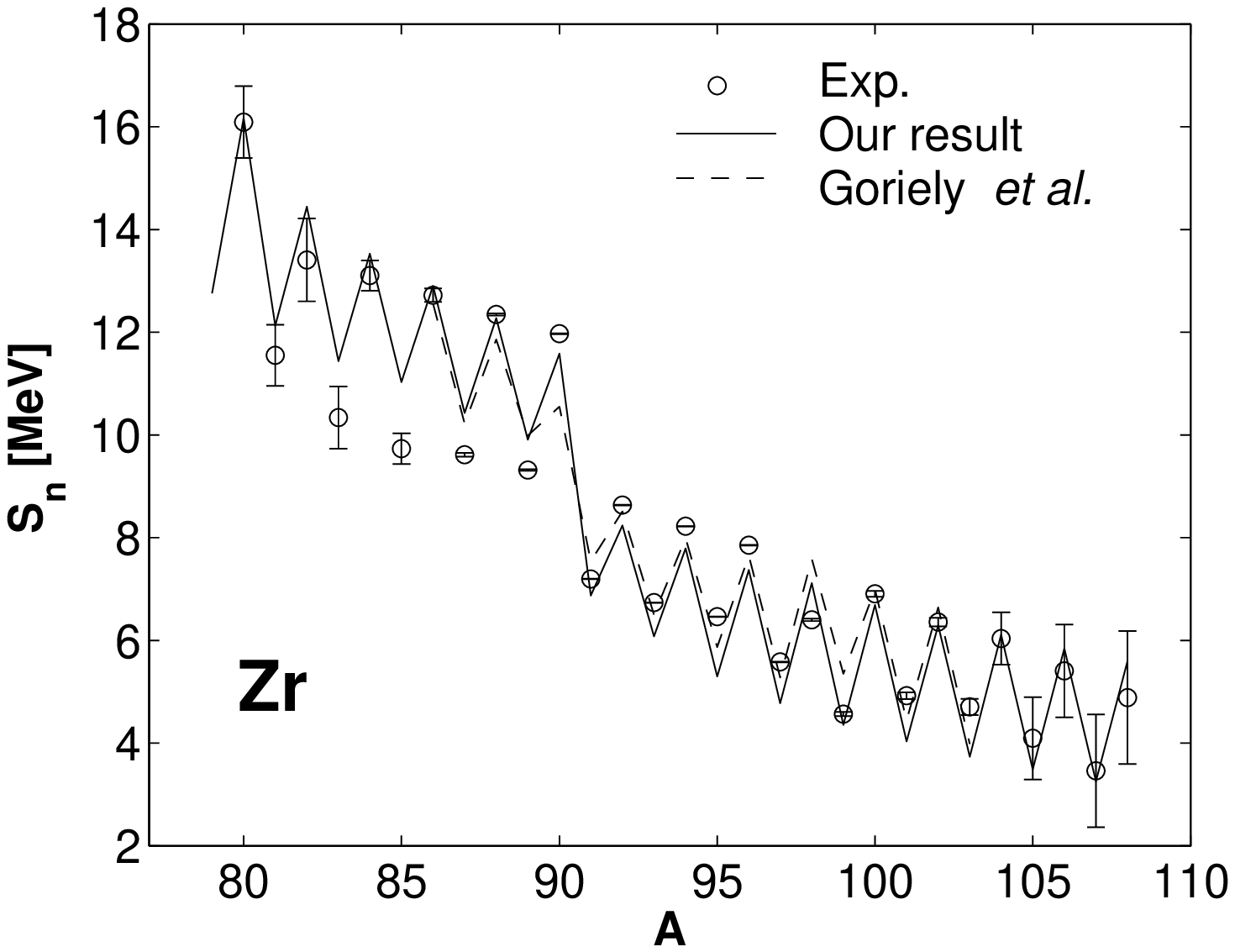}}
\end{center}
\caption{ One--neutron separation energies for zirconium
isotopes. Some zirconium isotopes are deformed according to Goriely
{\it et al.} }

\end{figure}

\begin{figure}[tbh]
\begin{center}
\epsfxsize=10.0cm
\centerline{\epsffile{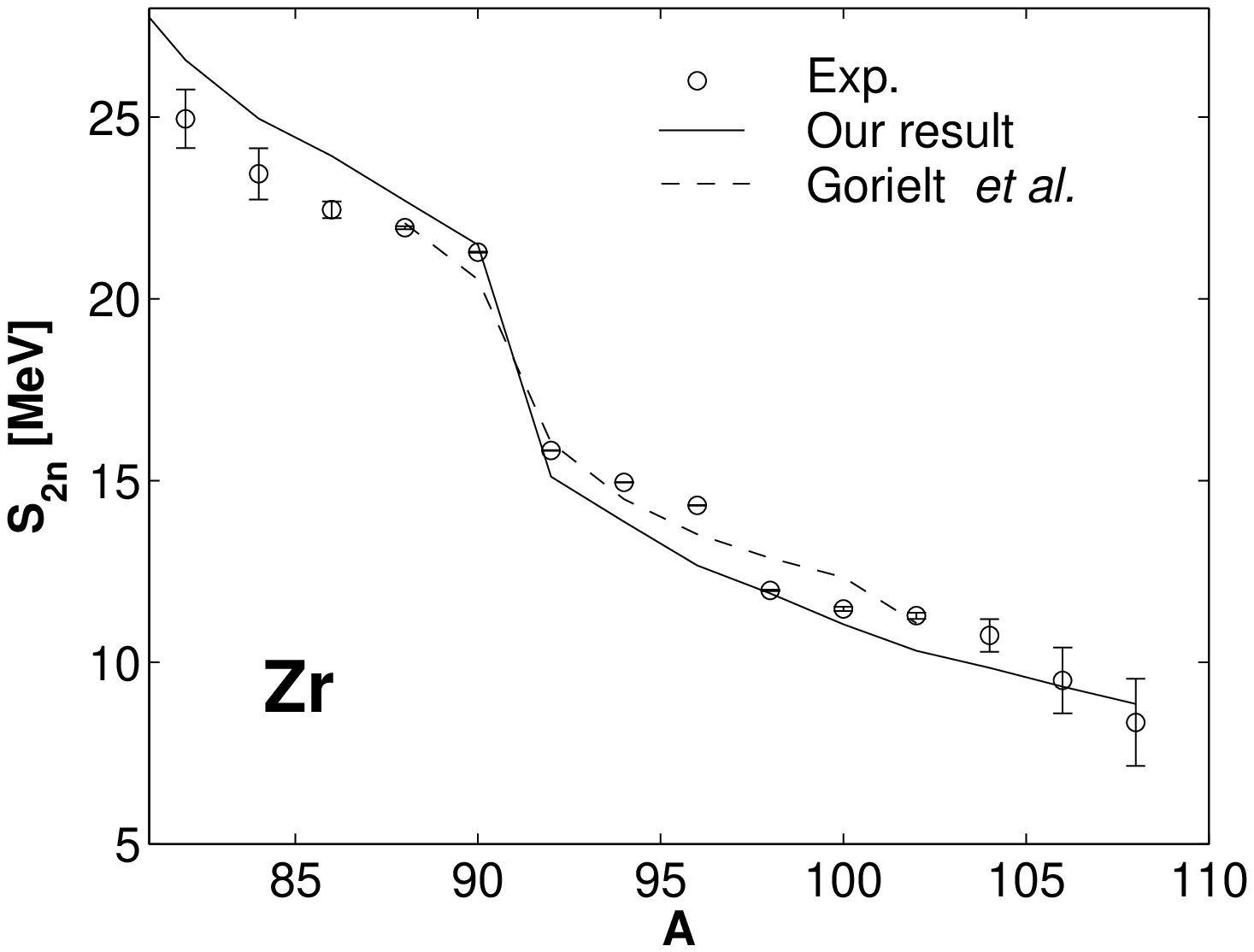}}
\end{center}
\caption{ Two--neutron separation energies for zirconium
isotopes. Some zirconium isotopes are deformed according to Goriely
{\it et al.}}
\end{figure}

\begin{figure}[tbh]
\begin{center}
\epsfxsize=10.0cm
\centerline{\epsffile{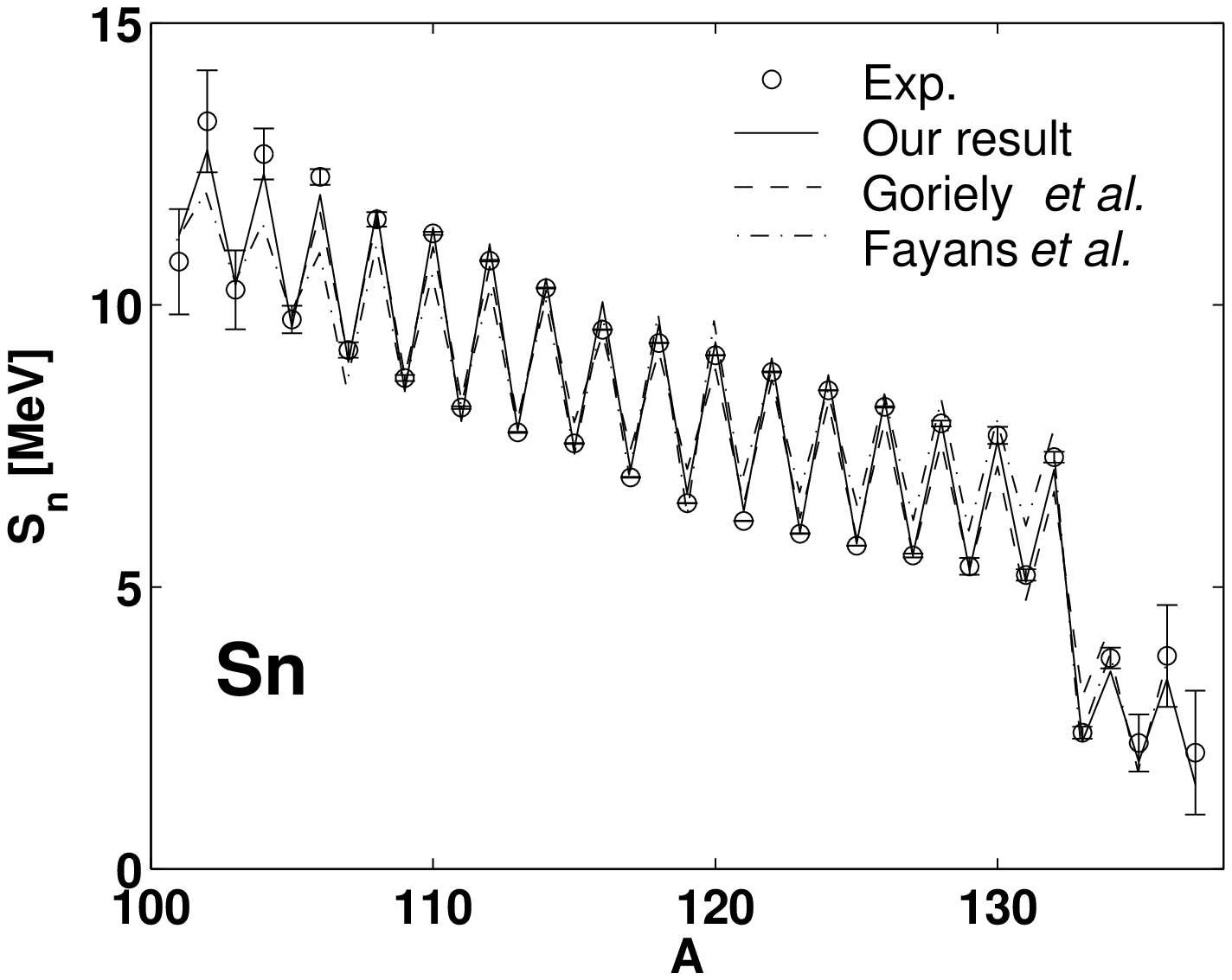}}
\end{center}
\caption{ One--neutron separation energies for tin isotopes.The
results of Fayans {\it et al.} presented here are for a renormalized
pairing strength with a factor of 1.05. }
\end{figure}

\begin{figure}[tbh]
\begin{center}
\epsfxsize=10.0cm
\centerline{\epsffile{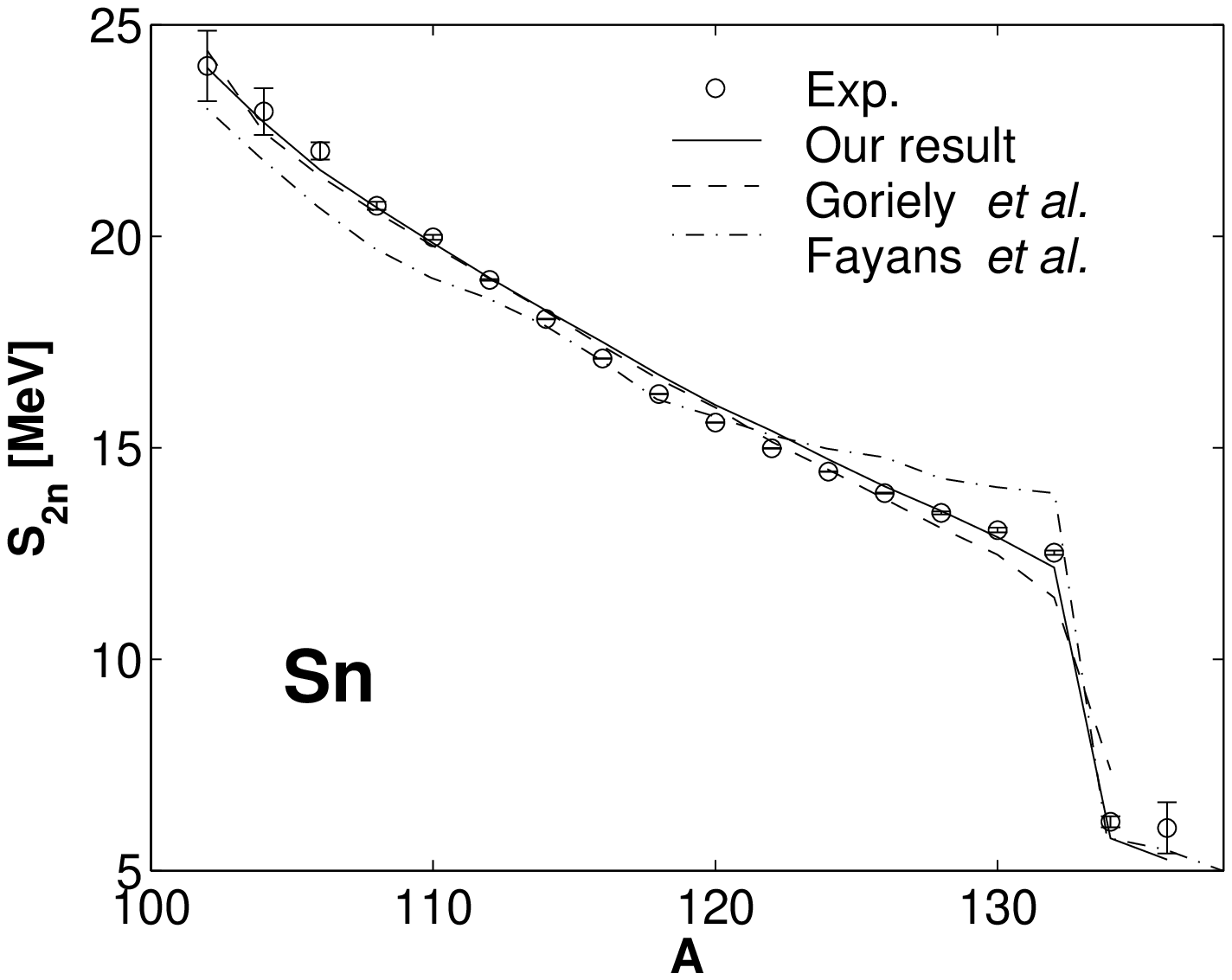}}
\end{center}
\caption{ Two--neutron separation energies for tin isotopes. The
results of Fayans {\it et al.} presented here are for a renormalized
pairing strength with a factor of 1.05.}
\end{figure}

\begin{figure}[tbh]
\begin{center}
\epsfxsize=10.0cm
\centerline{\epsffile{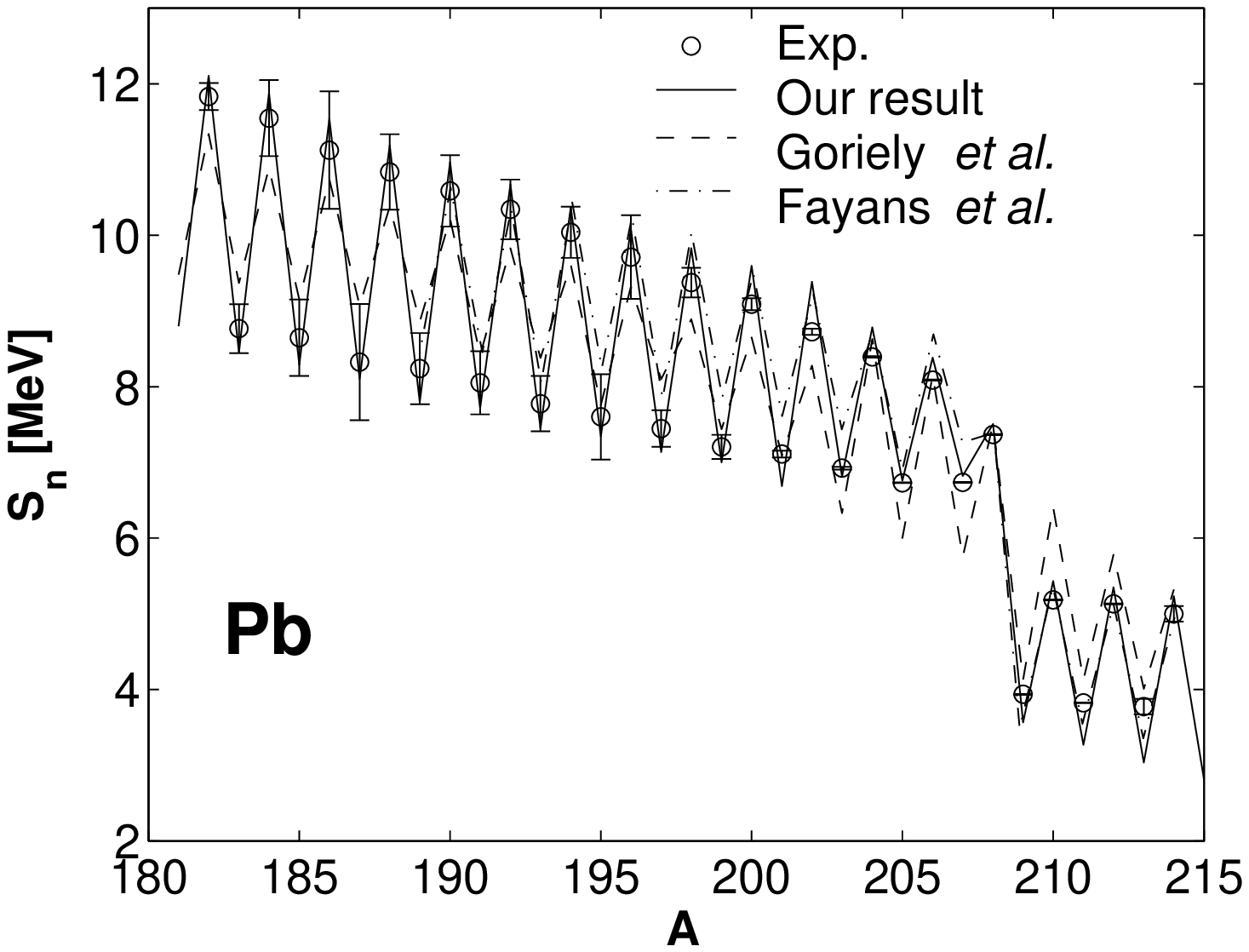}}
\end{center}
\caption{ One--neutron separation energies for lead isotopes.}
\end{figure}

\begin{figure}[tbh]
\begin{center}
\epsfxsize=10.0cm
\centerline{\epsffile{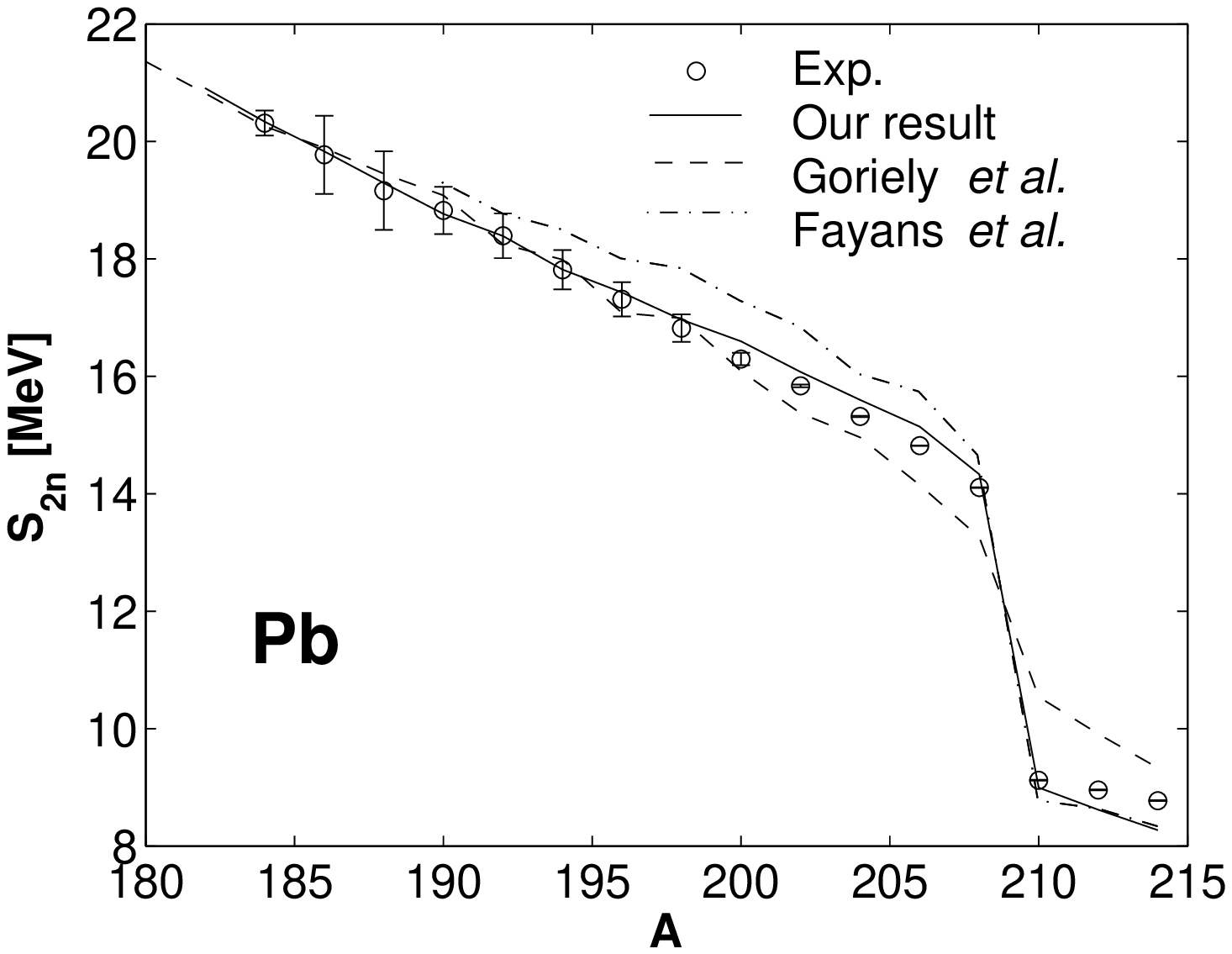}}
\end{center}
\caption{ Two--neutron separation energies for lead isotopes.}
\end{figure}

\begin{figure}[tbh]
\begin{center}
\epsfxsize=10.0cm
\centerline{\epsffile{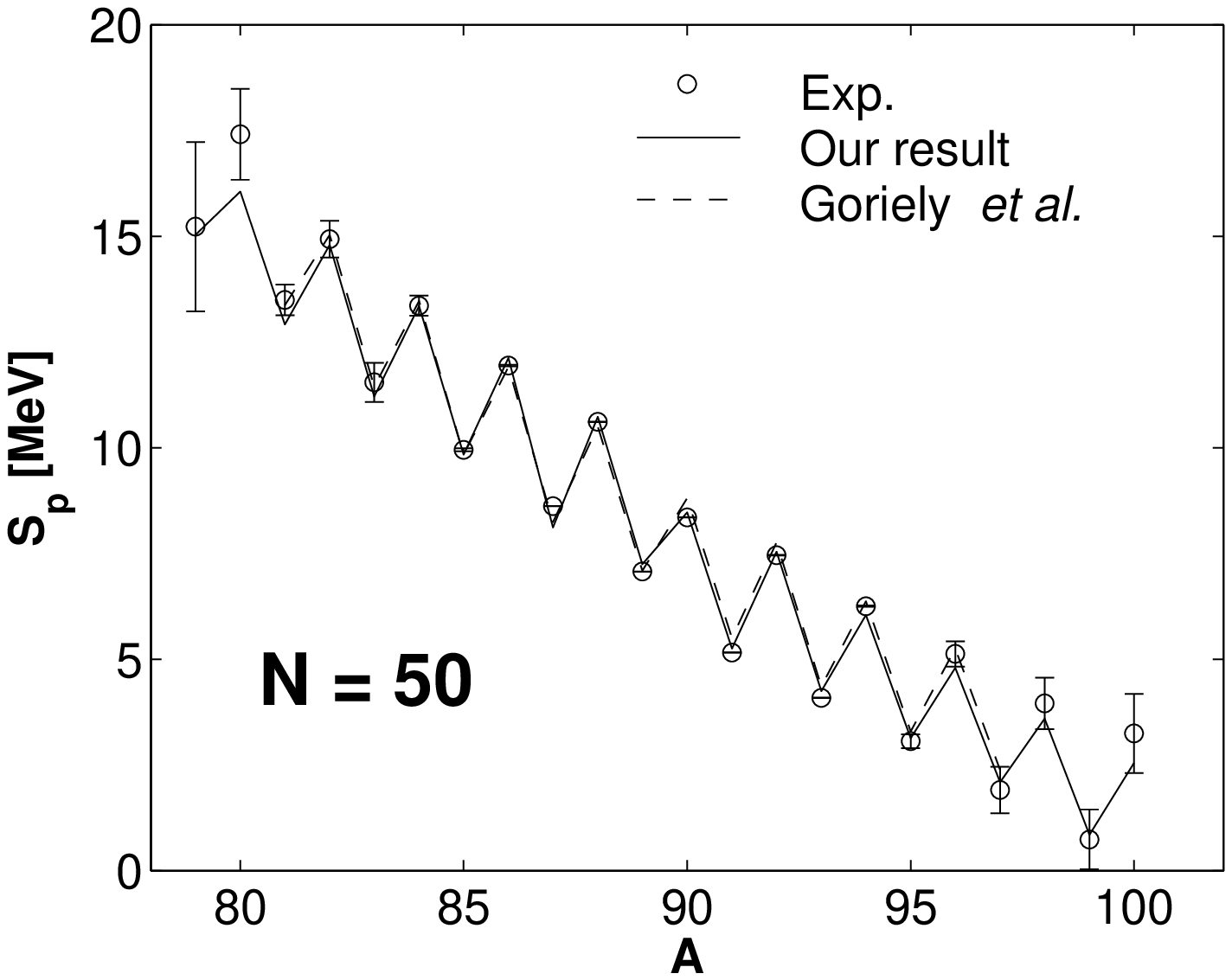}}
\end{center}
\caption{ One--protron separation energies for $N=50$ isotones.}
\end{figure}

\begin{figure}[tbh]
\begin{center}
\epsfxsize=10.0cm
\centerline{\epsffile{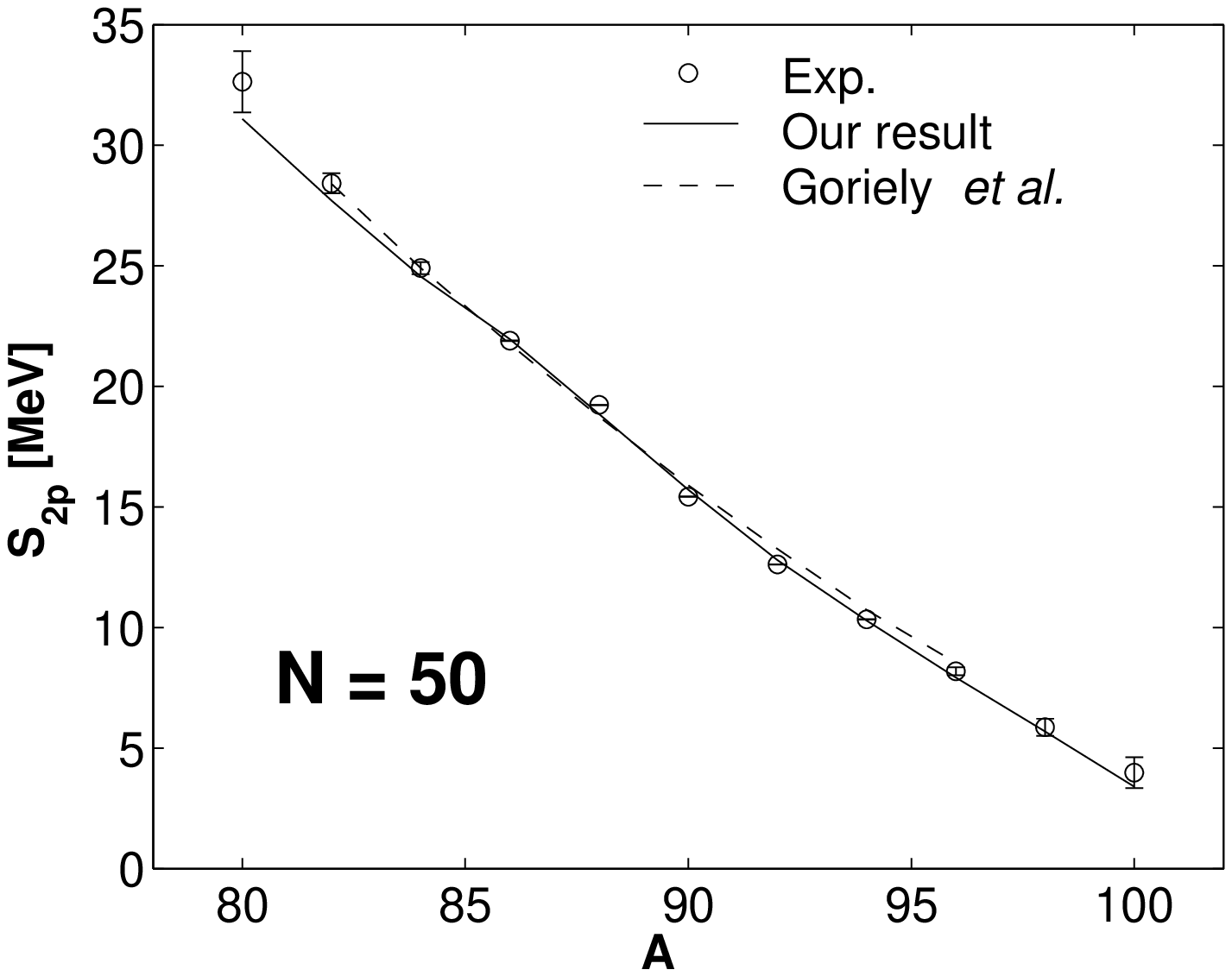}}
\end{center}
\caption{ Two--protron separation energies for $N=50$ isotones.}
\end{figure}

\begin{figure}[tbh]
\begin{center}
\epsfxsize=10.0cm
\centerline{\epsffile{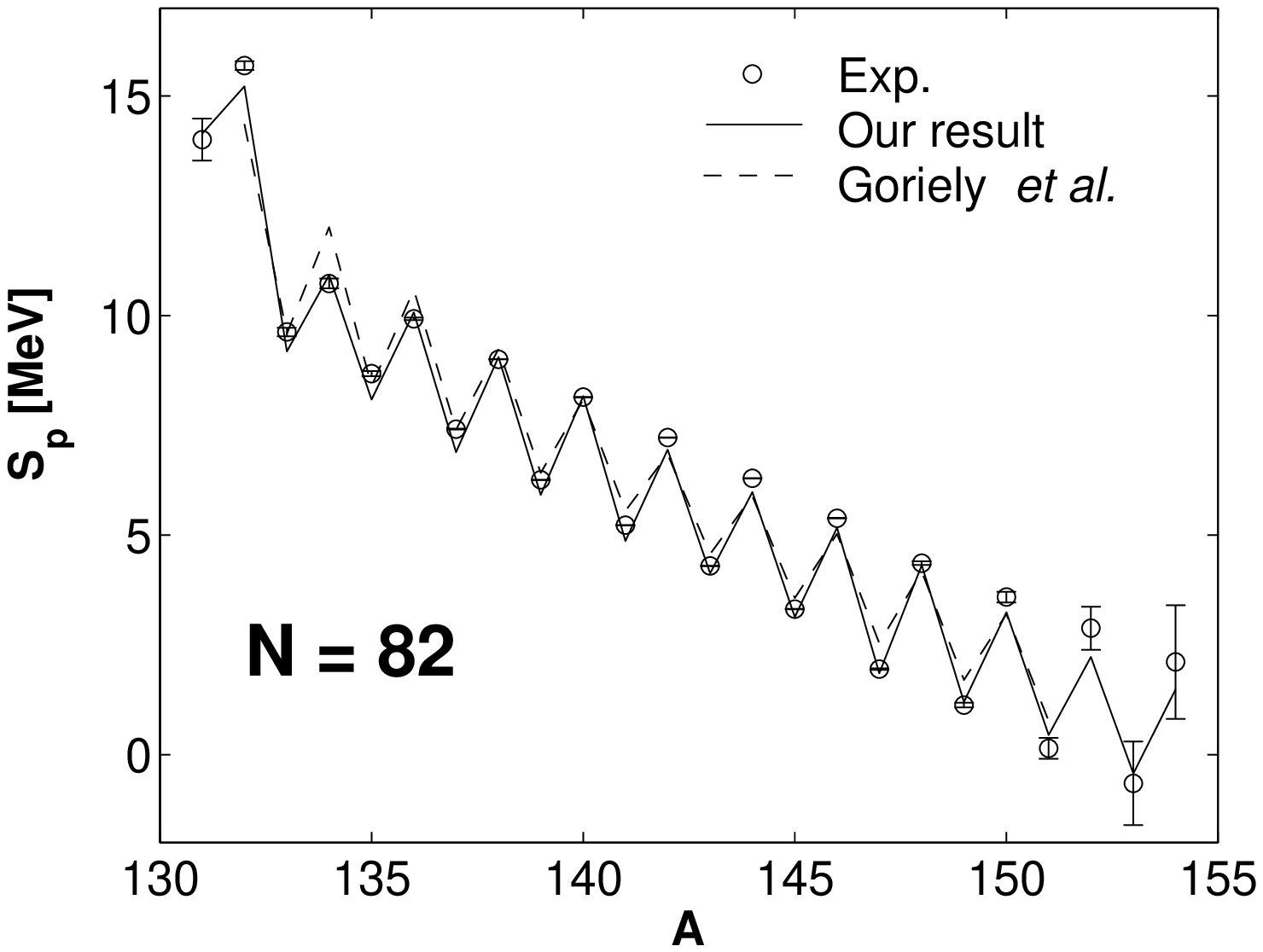}}
\end{center}
\caption{ One--protron separation energies for $N=82$ isotones.}
\end{figure}

\begin{figure}[tbh]
\begin{center}
\epsfxsize=10.0cm
\centerline{\epsffile{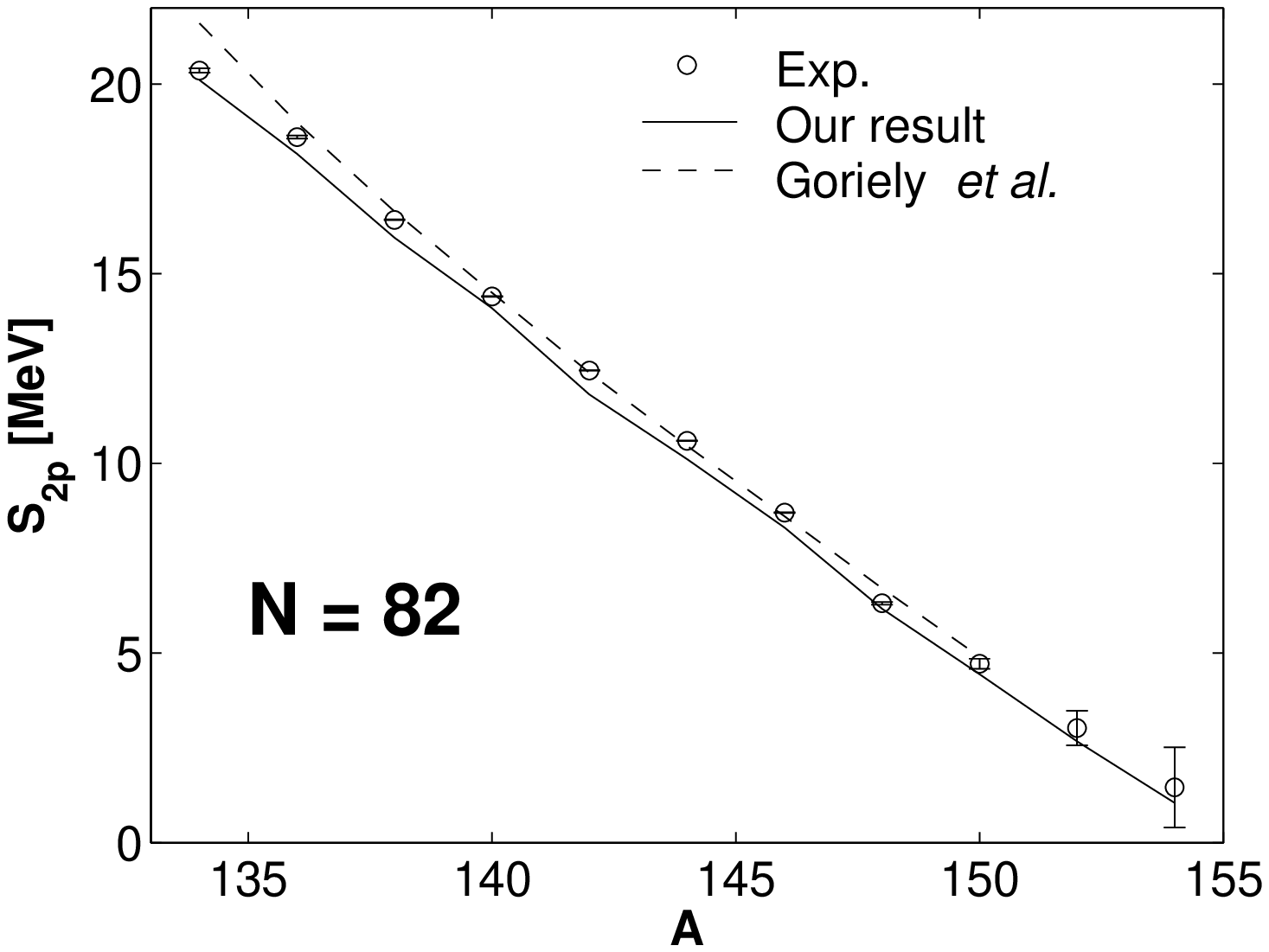}}
\end{center}
\caption{ Two--protron separation energies for $N=82$ isotones.}
\end{figure}

\begin{figure}[tbh]
\begin{center}
\epsfxsize=10.0cm
\centerline{\epsffile{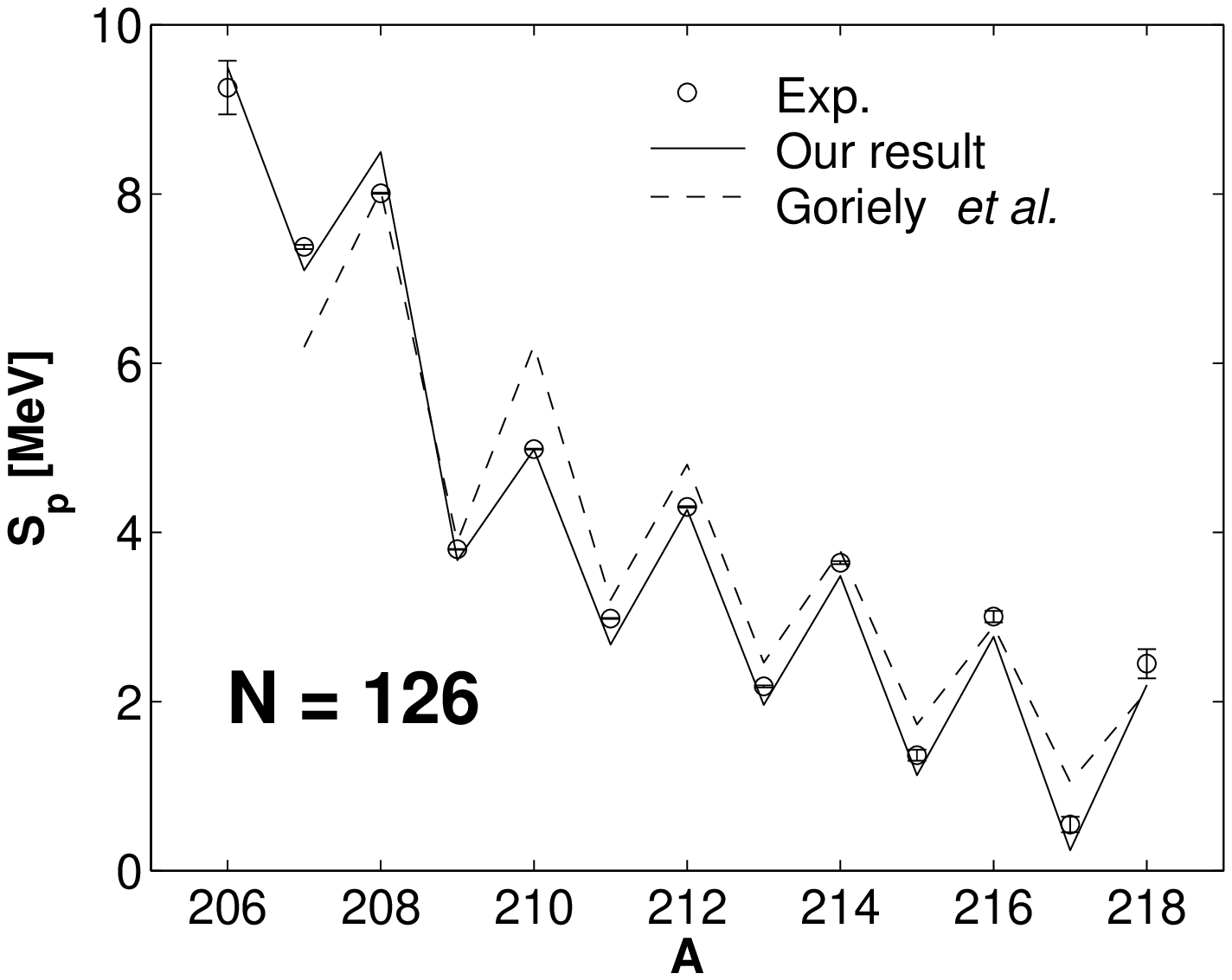}}
\end{center}
\caption{ One--protron separation energies for $N=126$ isotones.}
\end{figure}

\begin{figure}[tbh]
\begin{center}
\epsfxsize=10.0cm
\centerline{\epsffile{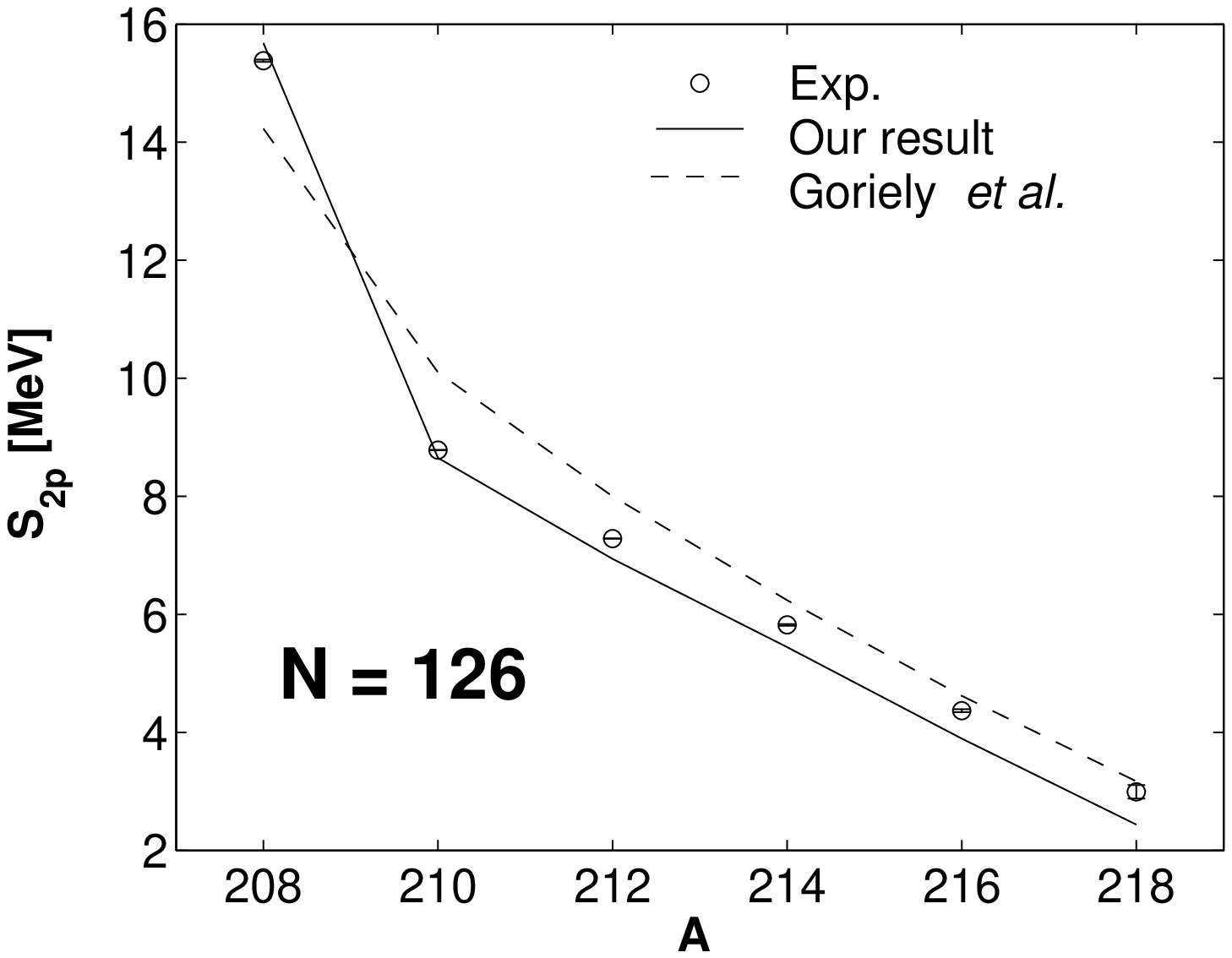}}
\end{center}
\caption{ Two--protron separation energies for $N=126$ isotones.}
\end{figure}

\begin{figure}
\begin{center}
\epsfxsize=10.0cm
\centerline{\epsffile{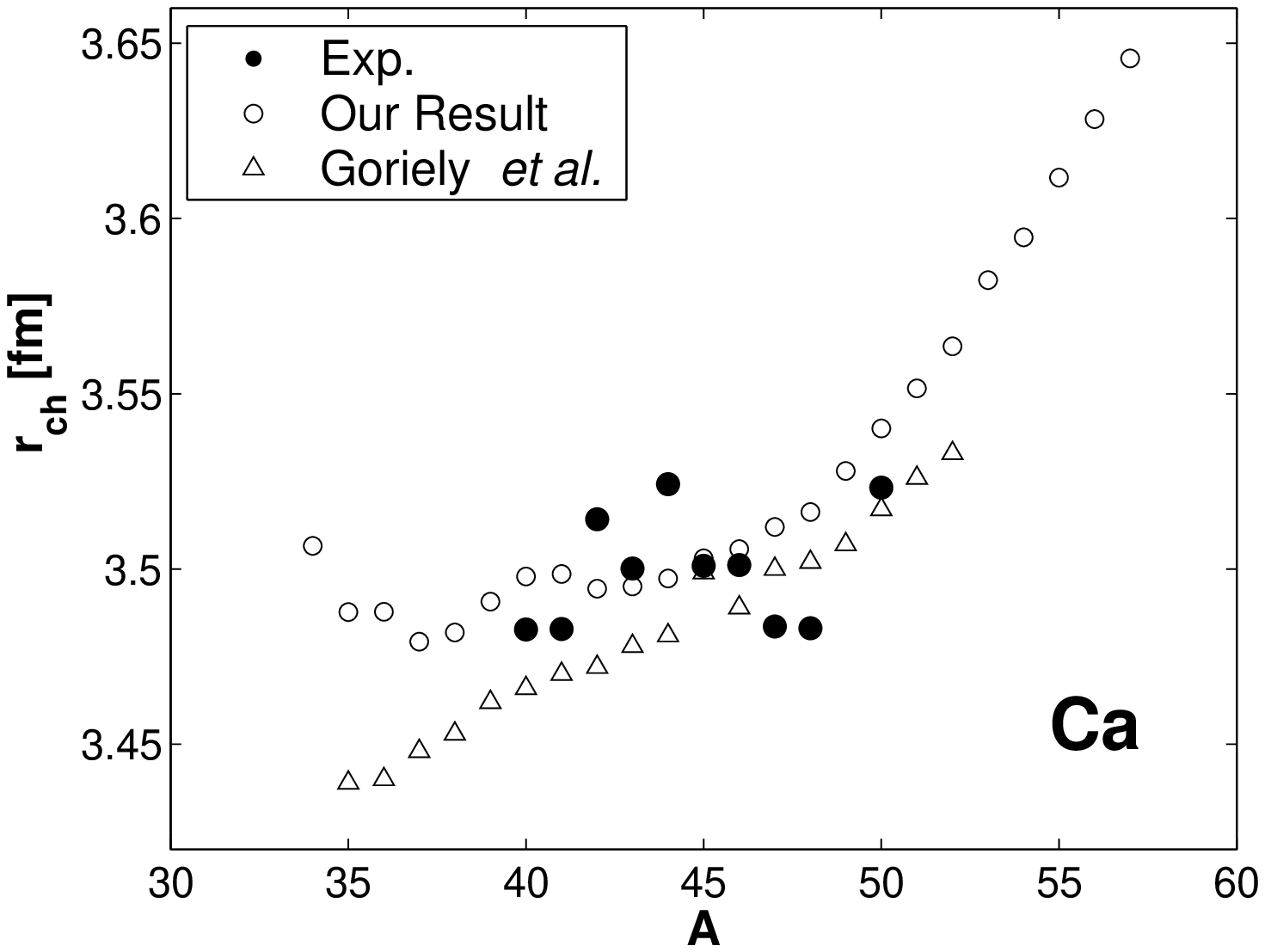}}
\end{center}
\caption{ Charge radii for calcium isotopes, our results (open
circles), Goriely {\it et al.} (open triangles) and experimental
values from Nadjakov {\it al et al.} (bullets).  }
\end{figure}

\begin{figure}
\begin{center}
\epsfxsize=10.0cm
\centerline{\epsffile{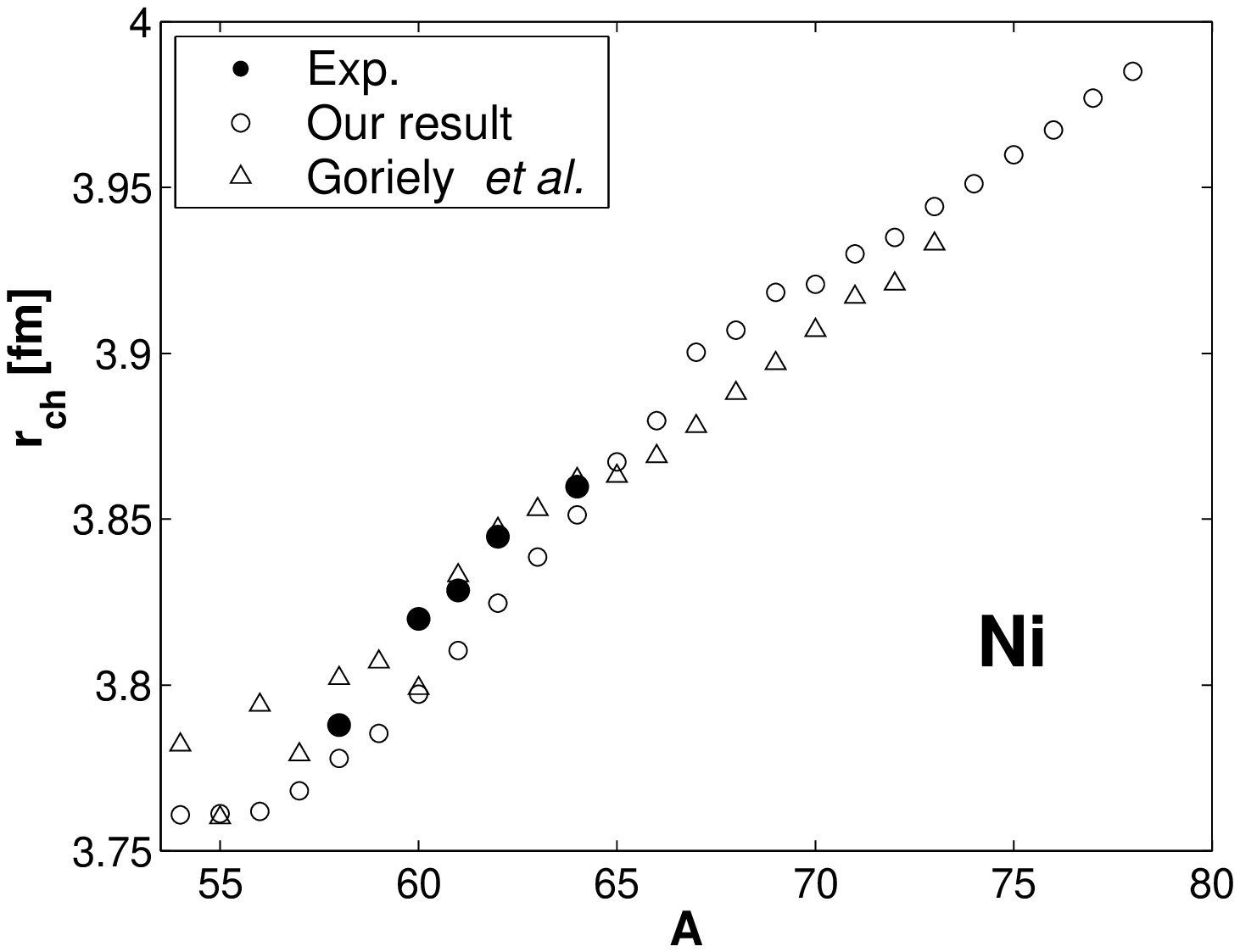}}
\end{center}
\caption{ Charge radii for nickel isotopes, our results (open
circles), Goriely {\it et al.} (open triangles) and experimental
values from Nadjakov {\it al et al.} (bullets). }
\end{figure}

\begin{figure}
\begin{center}
\epsfxsize=10.0cm
\centerline{\epsffile{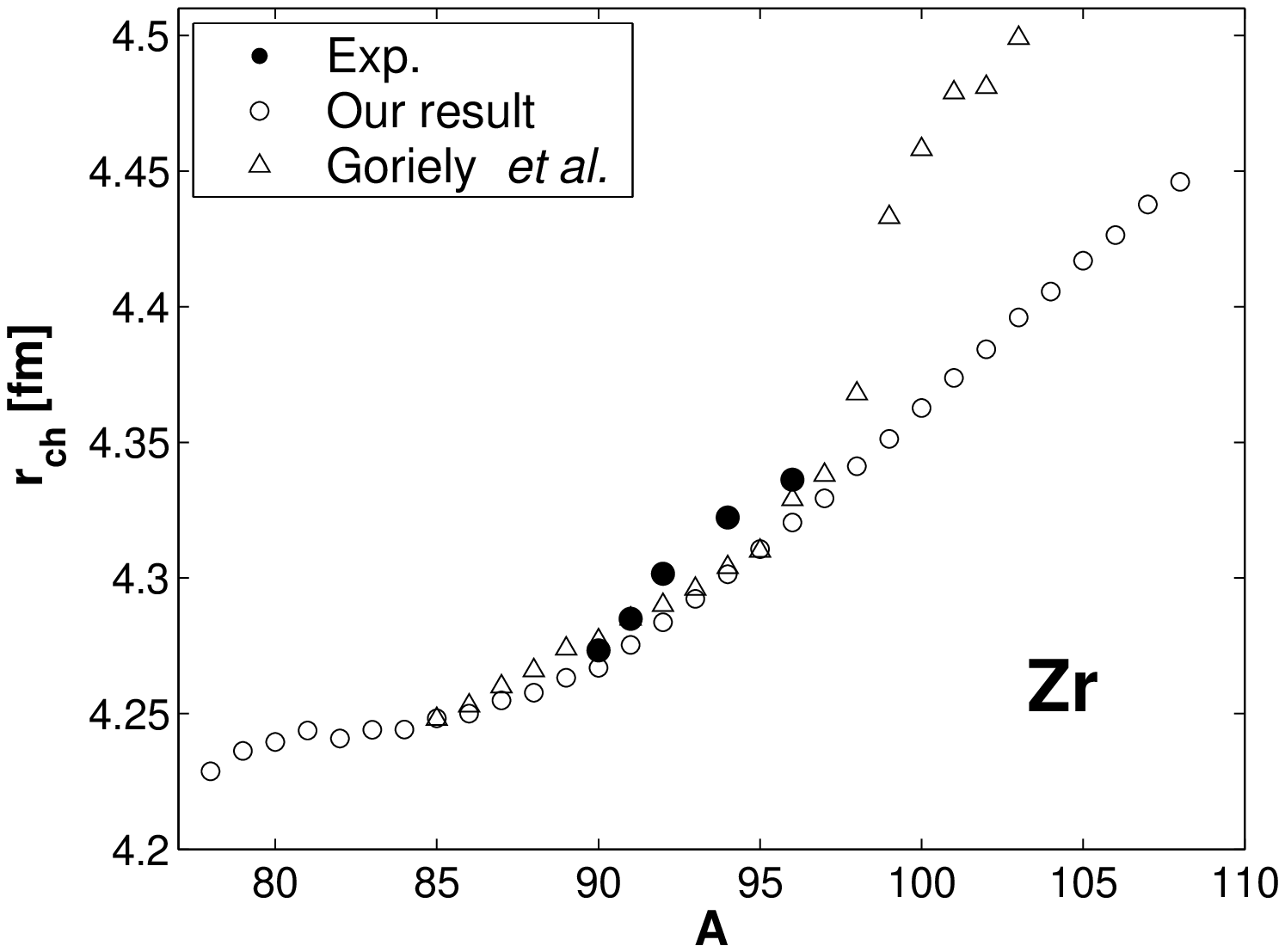}}
\end{center}
\caption{ Charge radii for zirconium isotopes, our results (open
circles), Goriely {\it et al.} (open triangles) and experimental
values from Nadjakov {\it al et al.} (bullets). }

\end{figure}

\begin{figure}
\begin{center}
\epsfxsize=10.0cm
\centerline{\epsffile{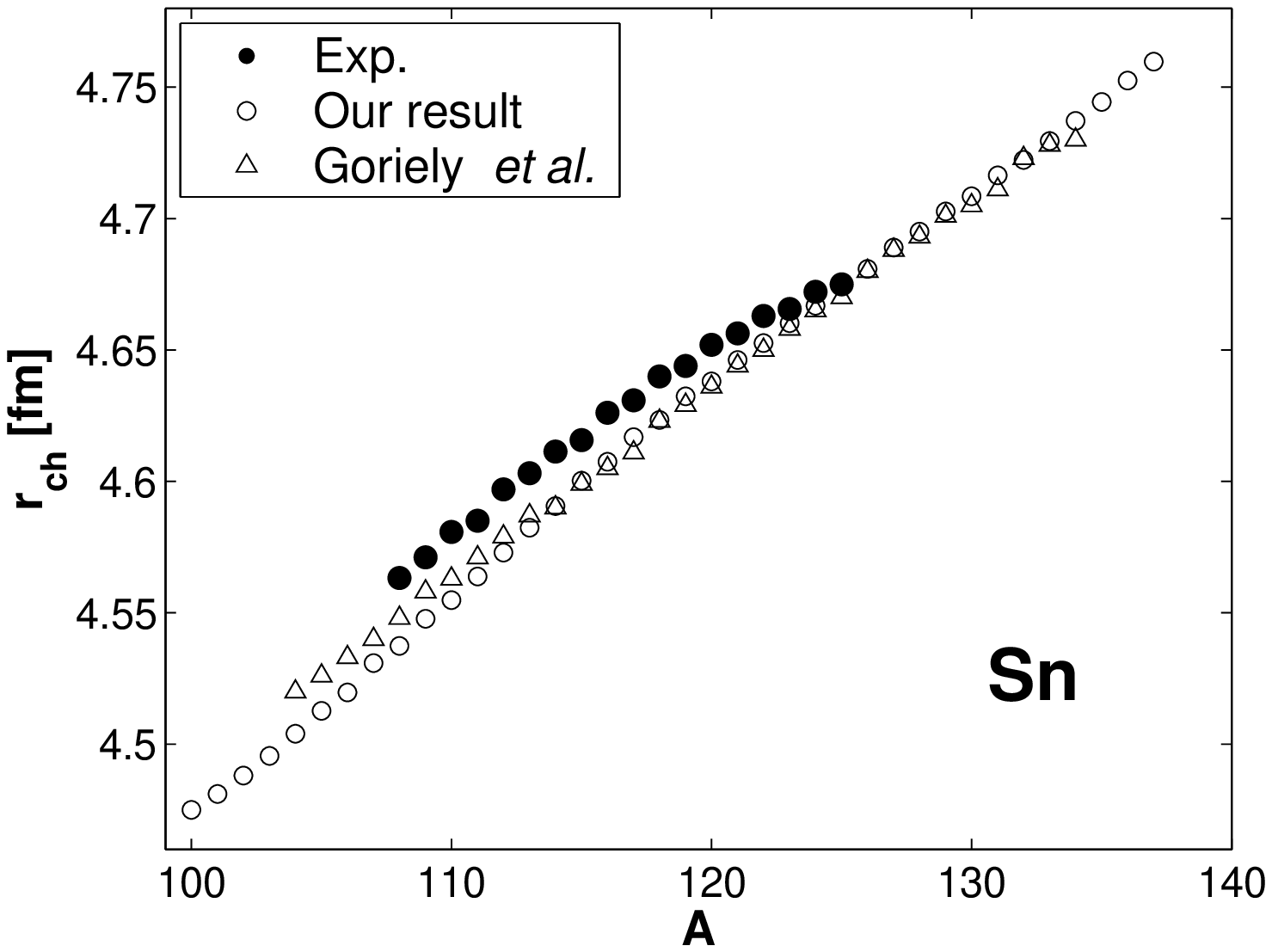}}
\end{center}
\caption{ Charge radii for tin isotopes, our results (open circles),
Goriely {\it et al.} (open triangles) and experimental values from
Nadjakov {\it al et al.} (bullets).}

\end{figure}

\begin{figure}
\begin{center}
\epsfxsize=10.0cm
\centerline{\epsffile{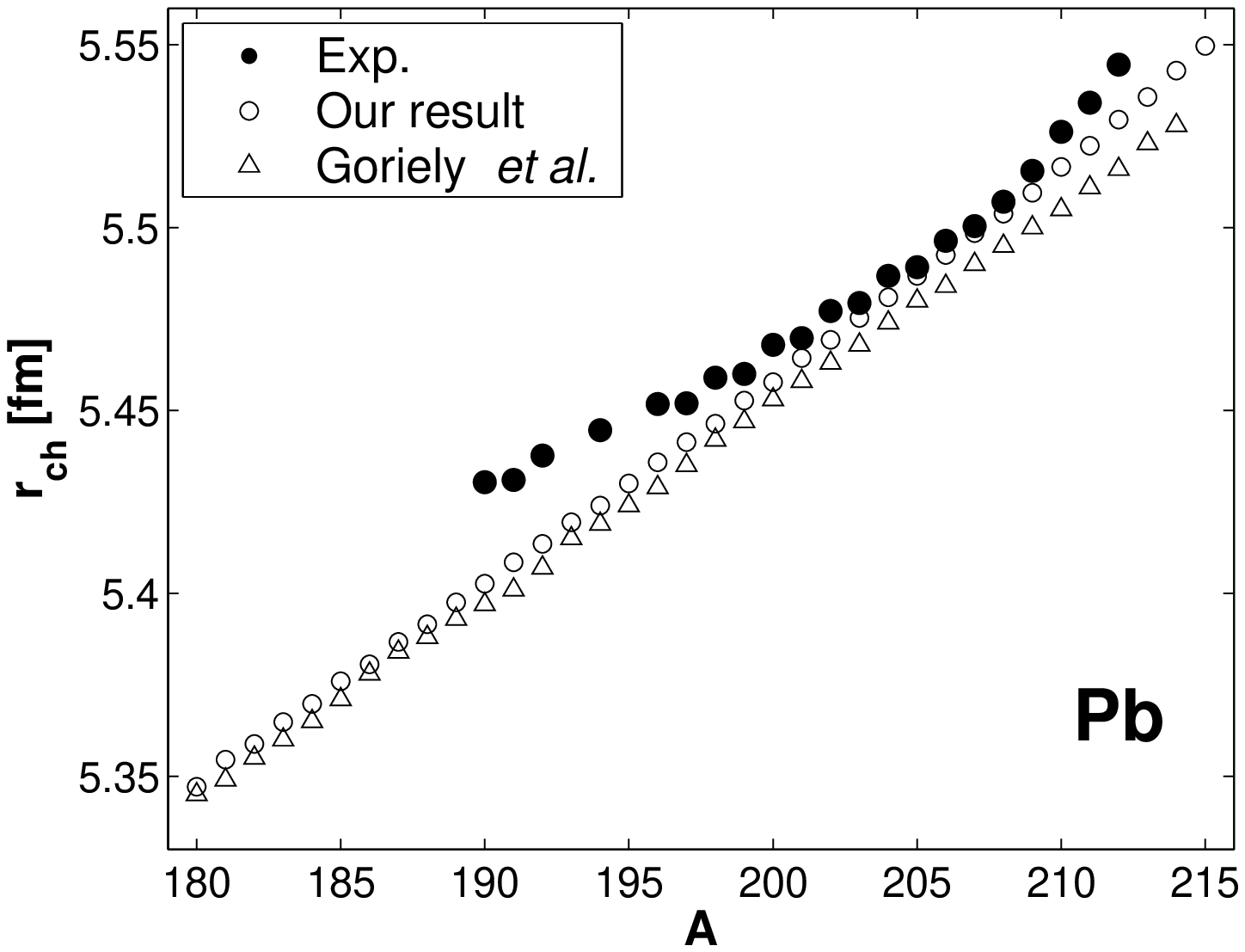}}
\end{center}
\caption{ Charge radii for lead isotopes, our results (open
circles), Goriely {\it et al.} (open triangles) and experimental
values from Nadjakov {\it al et al.} (bullets).}
\end{figure}


\end{document}